\documentclass{elsart}

 \usepackage{graphics}
 \usepackage{graphicx}

\usepackage{amssymb}

\usepackage{lineno}

\begin{document}
 \linenumbers

\begin{frontmatter}

\title{An Investigation on Cooling of CZT Co-Planar Grid Detectors}

\author{J. V. Dawson\corauthref{cor1}}
 \corauth[cor1]{corresponding author}
\ead{jaime.dawson@gmail.com}
\author{C. Montag, C. Reeve, J. R. Wilson, K. Zuber}
 \thanks[url]{COBRA collaboration: http://cobra.physik.uni-dortmund.de}
\address{Department of Physics and Astronomy, University of Sussex,
  Falmer, Brighton. BN1 9QH  UK }

\begin{abstract}
The effect of moderate cooling on CdZnTe semiconductor detectors has
been studied for the COBRA experiment. Improvements in energy resolution and low energy
threshold were observed and quantified as a function of temperature.
Leakage currents are found to contribute typically $\sim$5 keV to the
widths of photopeaks.  
\end{abstract}

\end{frontmatter}

\section{Introduction}
\label{intro}
In recent years a lot of effort has gone into the understanding of
new semiconductor materials.  There is much industrial and scientific interest in the development of
Cadmium Zinc Telluride(CZT) detectors because of their high stopping
power and room temperature operation.   Such devices have a wide field
of application in hard
X-ray/$\gamma$-ray astronomy, medical imaging applications and
general radiation detection like dosimetry.

A novel application is their usage to search for rare nuclear decays.
The COBRA experiment is planning to use a large array of Cadmium Zinc
Telluride (CZT) semiconductors, to search for neutrinoless double beta
decays \cite{zuber2001}.  Neutrinoless double beta decay, the
simultaneous decay of two neutrons inside a nucleus with the emission
of two electrons only is not allowed in the Standard Model of Particle
Physics, it requires that a neutrino is its own antiparticle as well
as that it has a non-vanishing rest mass.  For more details see
\cite{zuber2001}.  

For a double beta decay experiment, CZT offers great potential.  Of
the 35 known isotopes able to undergo double beta decay, CZT contains
5 of them as shown in Table \ref{table:Qvalues}.  Of special interest are $^{116}$Cd and $^{130}$Te due to their high Q values.

\begin{table}[]
\caption{Double Beta Decay Isotopes present in CdZnTe }
{\footnotesize
\begin{tabular}{@{}crrr@{}}
\hline
{Isotope} &{Q-value keV}\\
\hline
{$\beta\beta$ emitters} \\

{$^{114}$Cd} &{534}  \\
{$^{116}$Cd} &{2805} \\
{$^{128}$Te} &{868} \\
{$^{130}$Te} &{2529} \\
{$^{70}$Zn} &{1001} \\

{$\beta^{+}\beta^{+}$ emitters} \\
{$^{108}$Cd} &{231} \\
{$^{106}$Cd}&{2771}\\
{$^{120}$Te}&{1722}\\
{$^{64}$Zn} &{1096} \\
\end{tabular}\label{table:Qvalues} }
\vspace*{-13pt}
\end{table}

Such decays would produce signals due to the combined energy deposit of the two emitted electrons. Since no neutrinos are emitted in this process, a peak is observed (broadened by the energy resolution of the detectors) at the Q-value of each decay.  The rate of these events is proportional to the mass of the neutrino.

For decays in which neutrinos are emitted, the allowed double beta decay process, the observed total energy deposits like below the Q-value, forming a spectrum similar to that observed from beta decay.  The experimental challenge is to be able to separate the peak due to the neutrinoless mode from the spectrum produced by the more frequent allowed mode.  This requires a good energy resolution, typically less than 2\% at 2.8 MeV.

Additionally 4 of the isotopes can also undergo double positron decay, hence emitting positrons instead of electrons ($\beta^{+}\beta^{+}$)as shown in Table \ref{table:Qvalues}.
Although, due to energy constraints only $^{106}$Cd can emit two
positrons, whilst the other isotopes decay via mixed modes of positron
and single electron
capture of a K-shell electron ($\beta^{+}/EC$) and double electron
capture ($EC/EC$) modes.  

For the modes which emit one or more positrons, the resultant signal can also comprise one or more 511 keV annihilation photons. In a large array, it is possible that the 511 keV photons will be detected a CdZnTe crystal other than the one in which the decay occurred.  As each crystal signal is read-out separately this would be produce two or more coincident triggers, with known energy values in each crystal.  These coincident signals are characteristic of these processes and a powerful search technique.

The energy range of the observed events span a wide range from the detection
of X-rays only from double electron capture (around 60 keV for
$^{106}$Cd decay) up to a pair of electrons with total energy 2.8 MeV for $^{116}$Cd double beta decay.

Common to all these decays is the fact that they are very rare, with
half-lives well beyond 10$^{20}$ years.  For the latest COBRA results
see \cite{Bloxham2007}.  The half-life sensitivity of
the experiment is highly dependent on the energy resolution of the
detectors. For the background limiting case, in which one observes no
events, the lower limit on the half-life $ T^{1/2}$ comes from the Poisson fluctuation on the total number of background counts observed within the peak region.  The lower limit on the half-life $ T^{1/2}$ (years) is:
  \begin{equation}
    T^{1/2} = ln(2). \epsilon .N_{A}. \frac{M}{M_{A}}  \sqrt{\frac{M.
    t}{\Delta E . B}} \end{equation}
where $\epsilon$ is the efficiency, $N_{A}$ is Avagadro's number, $M$ is
    the mass of isotope in kg, $M_{A}$ is the atomic mass (kg), $t$ is the measuring
    time (years), $B$ is the background rate (events
    kg$^{-1}$keV$^{-1}$year$^{-1}$) and $\Delta E$ is the energy
    resolution (keV).  
  
For a full-scale experiment, with which one would be sensitive to a neutrino mass of $\sim$50 meV, the mass of isotope required is typically 100 kg. In such an experiment, the rate of neutrinoless double beta decays observed may be less than a few per year.  

To detect such low count rates requires a detector and surrounding components of extremely high radiopurity.  Shielding from cosmic rays and ambient trace radioactive backgrounds is essential.  Experiments are sited underground, for COBRA in Laboratori Nationali del Gran Sasso (LNGS, Italy).  The local radioactivity is combatted using thick neutron shields and lead castles.  Detectors are usually housed inside an inner ultra-clean copper shield which protects against the trace radioactivity of the lead castle. One possible large source of radioactive contamination comes from the detector electronics.  For the COBRA experiment, like other semiconductor rare search experiments, the pre-amplifier electronics are placed outside the lead castle, well separated from the inner crystals. This has one obvious drawback, that the detector energy resolution is thus compromised and effectively degraded due to the long signal cable lengths needed.

Other physics searches are also possible such as the second-forbidden unique electron capture of $^{123}$Te \cite{Munstermann2002, Alessandrello2003}, which would produce a peak at 30.5 keV.  The group has also performed measurements of the four-fold forbidden non-unique beta decay of $^{113}$Cd \cite{Goessling_2005} which has a Q-value of 320$\pm$2 keV \cite{tableofisotopes}.  The shape of the beta decay spectrum is not well predicted by theory, and has only been observed by a few experiments. 

To maximise the physics output of the COBRA experiment the ideal energy range observable by each subcomponent detector is between 25 keV and 3 MeV, with the best possible energy resolution.

The current COBRA incarnation is a small R\&D array, designed with the aim of testing the coincident search technqiue.  It comprises 64 CdZnTe semiconductors of 1 cm$^{3}$ organised in a 4$\times$4$\times$4 array.  The total mass of the experiment is $\sim$ 400g.  The crystal support structure is manufactured from delrin, and placed in a copper shield of 15 cm thick.  The surrounding lead shield is 20 cm thick.  A narrow slit in the copper shield and a V-shaped lead brick allows the detector cabling to pass through the shielding layers to the pre-amplifier electronics.  

\subsection{The Detectors}
\label{intro_det}

The COBRA crystals were supplied by EV PRODUCTs, and due to the
quantity required and financial constraints, are all low-grade.  The
crystals were supplied without contacts; ie no wires glued to the
electrodes, as all contacting methods commercially used are not sufficiently radiopure
for a low background experiment. Typical methods of contacting are
gluing with conductive epoxy or soldering.  The crystals supplied are
exactly the same in electrode design and size as those purchased
pre-contacted and housed.
 
Techniques are sought to improve the energy resolution of this
experiment whilst maintaining the low background.  In this paper we
consider the effect of slight cooling of the CZT detectors which could
be provided by delivering cool nitrogen gas, from liquid nitrogen boil
off, directly into the heart of the experiment.  This would be a
convenient technique since warm nitrogen gas is already delivered into
the crystal housing in order to flush out radon gas. 

The aim of this paper is to explore the effect of slight
cooling of CZT detectors on energy resolution in general and the
improvement for double beta searches specifically.

This work presents a systematic study of the effect of temperature on three Co-Planar Grid (CPG) crystals all with size 1x1x1 cm$^{3}$ and all manufactured by eV PRODUCTS, of which two were purchased uncontacted, the COBRA detectors, and one which was bought commercially.

The CPG structure forms a virtual Frisch grid below
the anode, and the resulting signal is produced by electrons travelling past
this virtual grid to the anode.  In this way there is almost no position
dependence on the signal amplitude and only the electron signal will be
readout \cite{luke1995}.

It is well known that slight cooling can dramatically reduce leakage
currents, and potentially enhance the observed energy resolution.  
Temperature effects on similar sized CPG detectors have previously
been reported \cite{Sturm2005,Amman2006}, however in these study the
preamplifier electronics were also cooled.  Effects on pixel
detectors has also been reported \cite{Yadav2005}.

\section{Experimental Setup}
\label{setup}
\begin{figure}
\includegraphics[width=9 cm, angle=0]{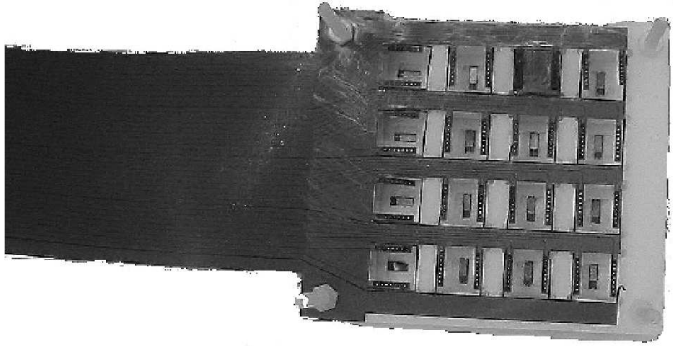}
\caption{The COBRA setup with 1 crystal.  16 crystals can be placed in to
  the Delrin  holder.  HV Kapton cable contacts to the cathode, and
  signal Kapton cable contacts the anodes.}
\label{fig:COBRAsetup}
\end{figure}

For a systematic study of CZT crystal response under slight
cooling, two 1x1x1 cm$^{3}$ CZT crystals with gold CPGs produced by eV
PRODUCTS were used.  These crystals, designated A and B, were purchased for the COBRA
experiment and setup as follows.  The two COBRA CPG crystals are seated in a delrin holder, with their
contacts bonded on to two Kapton cables using a homemade low radioactivity
conductive glue.  One Kapton cable supplies the HV to the cathodes,
and one is for the two anode connections (see photograph Figure
\ref{fig:COBRAsetup}).  The guard rings were not connected. This is
the usual manner to run the detectors for the COBRA experiment as it
keeps the active volume large.  The signal cable is $\sim$ 30 cm long.
This cable plugs in to COBRA designed preamplifier and subtraction circuit electronics, based on the recommendation of eV PRODUCTS  

All preamplified signals are then shaped by an Ortec 855 shaping
amplifier with a shaping time of 1$\mu$s, digitised by 2048-channel MCA card, and recorded by
computer.  The resulting spectra are analysed using ROOT peak finding
and continuum subtracting algorithms (TSpectrum), and the identified
 photopeaks fitted with a two-sided Gaussian (with different rising
 and falling sigmas). The Full Width Half Maximum is calculated using
 the average sigma.  A full sweep of all parameters; CPG balance potentiometer, grid bias
and cathode voltage, was made for both crystals with $^{137}$Cs
spectra recorded for each combination.  The shaping time was not optimised and remained at 1 $\mu$s.  The optimum parameters for each
crystal are those settings at which the best energy resolution of
the 662 keV photopeak was seen at room temperature.

The COBRA crystal holder was placed in a temperature controlled copper box, with a narrow slit to allow the Kapton cables to pass through. The copper box acts as a Faraday cage. An external peltier cooler is in close thermal contact with the exterior of the box. The thick copper walls (3 mm) produce a even thermal bath which is monitored by two temperature sensors inside.  The peltier itself is cooled by a closed-loop pumped water system and cooling fans.  The fan system produces a significant amount of noise which degrades the energy resolution of the detectors.  Since the copper box is significantly massive, the fan system is switched off during the measurement and the temperature monitored to ensure stability during the measurement.  The measurements are therefore relatively short, of 100 s duration, but are repeated to ensure repeatability.  The temperature range of interest is from
2$^{\circ}$C to 20$^{\circ}$C.

\section{Resolution Function of a Commercial Detector}
\label{resolutionfunctions}
All CPG detectors show a linear increase in Full Width Half Maximum with increasing energy.  A possible explanation for this trend could be from the coplanar grid. Whilst the CPG design counteracts the effect of the hole-trapping in CdZnTe, the electrode design may also limit the resolution \cite{He2005}.  An x-y scan was made across the
cathode surface of a commercial eV PRODUCTS detector with a collimated
60 keV source.  The detector was a 1 cm$^{3}$ detectors, bonded to eV PRODUCTS preamplifier and subtraction circuit, and housed in a cylindrical aluminium casing.  We observed a systematic linear change in the photopeak position of the 60 keV line along one axis of $\sim$ 5\%. This effect was first reported in \cite{He2005}.  Illuminating the whole cathode of the detector results in a wider photopeak due to this effect.

Another important source of resolution broadening is the
fluctuation of the leakage current.  For these detectors, however, we
find that leakage current is only significant for low energy events
(below a few hundred keV).

\begin{figure}
\begin{center}
\includegraphics[width=9cm, angle=0]{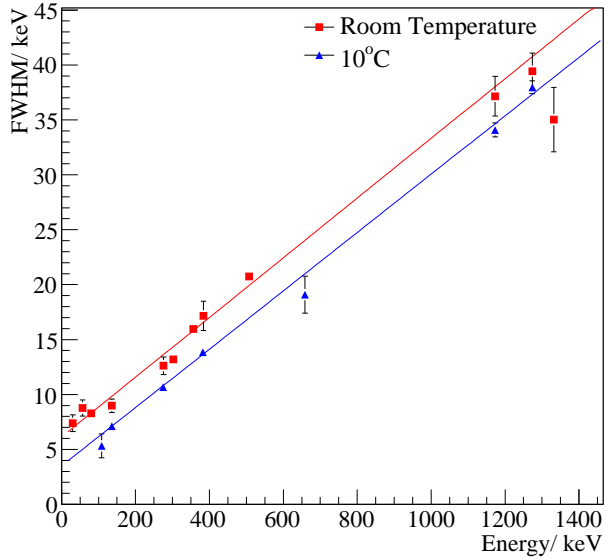}
\caption{Resolution Function for commercial detector at 22$^{\circ}$C and 10$^{\circ}$C}
\label{fig:resolutionfunction}
\end{center}
\end{figure}

A clear improvement was observed by cooling the commercial eV PRODUCTS detector as shown in
Figure \ref{fig:resolutionfunction}.  This clearly shows the linear
dependence of the FWHM with energy.  Results are shown for
measurements made at  room temperature for incident gamma rays and with the entire device
(crystal and pre-amplifier electronics) cooled to 10$^{\circ}$C.
Cooling this detector results in a 15\% improvement at 500 keV, and a
5\% improvement at 2.8 MeV. The shift downwards in intercept and
overall performance improvement is interpreted as due to the
reduction of bulk, surface leakage currents and electronic noise. 

\section{COBRA CZT Crystals}
After the encouraging results from the commercial detector the COBRA
 crystals were cooled, with the preamplifier electronics
remaining at room temperature (23$^{\circ}$C). No spectral change was observed for the photopeaks of $^{22}$Na (511 and 1254 keV),  $^{137}$Cs (662 keV) and $^{60}$Co (1173 and 1332 keV). However, improvements were observed with $^{241}$Am (60 keV) and $^{57}$Co (122 keV).

\subsection{Low Energy Response}
\label{lowenergy}
\begin{figure}
  \begin{minipage}{0.45\textwidth}
 \includegraphics[width=5.5cm,angle=-90]{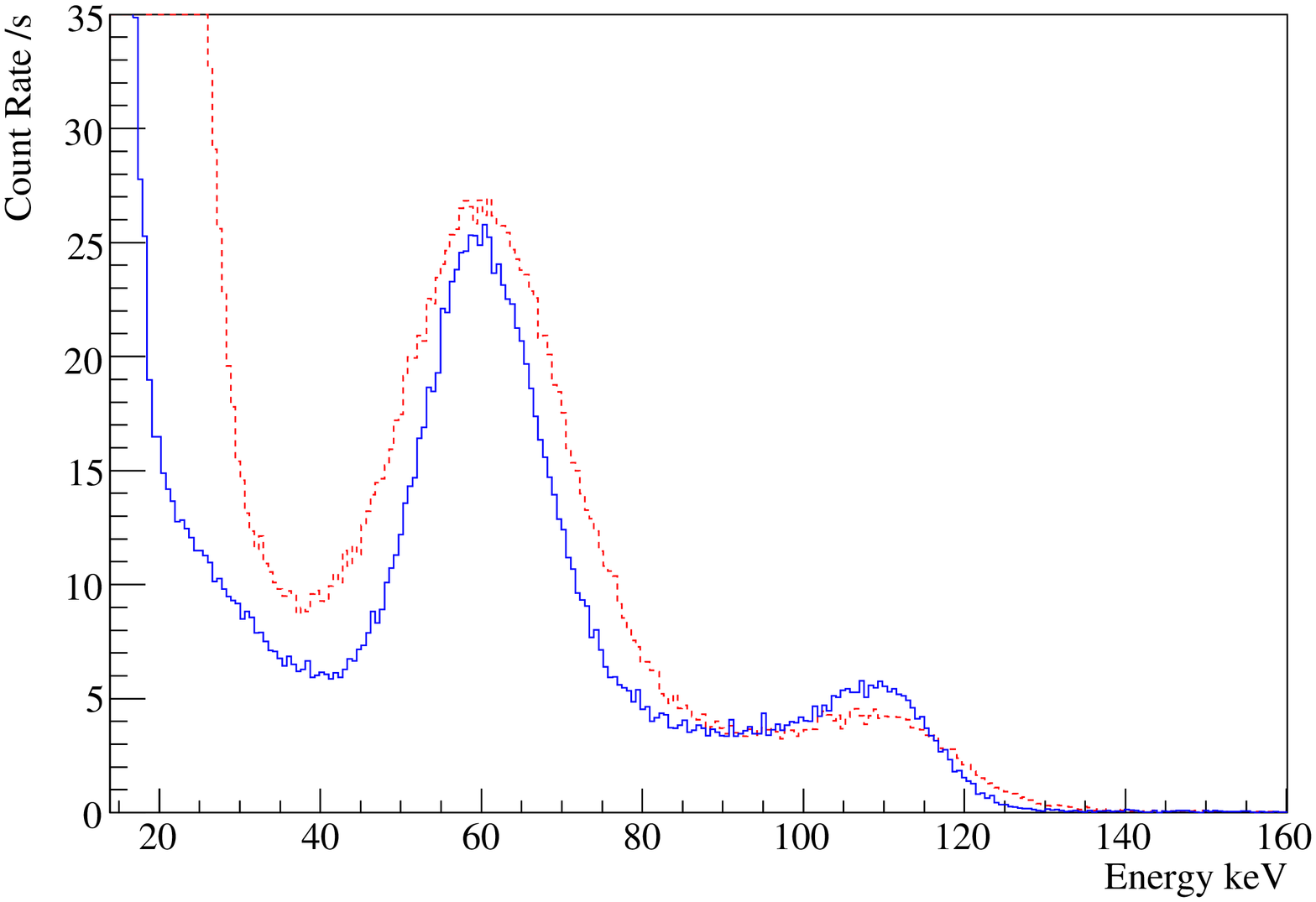}
  \end{minipage}%
 \hspace{0.1cm}
  \begin{minipage}{0.45\textwidth}
\includegraphics[width=5.5cm,angle=-90]{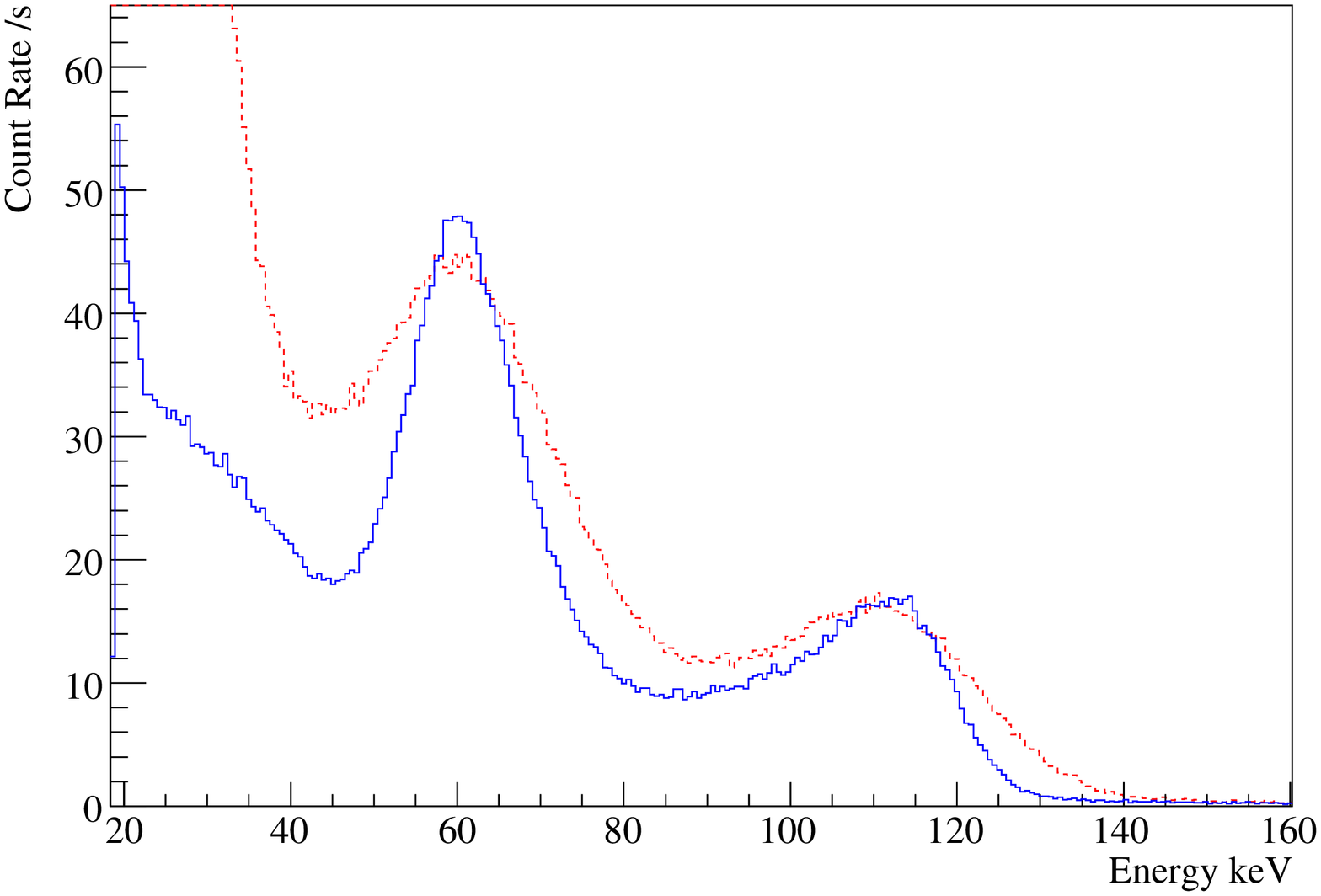}  \end{minipage}%
\caption{\label{fig:am241} $^{241}$Am spectrum of crystal A (left) and
B (right) at room temperature (23$^{\circ}$C) (dashed line) and under cooling (5$^{\circ}$C)(solid line).  The feature to the right of the 60 keV peak is due to multiple events.}
\end{figure}

\begin{figure}
  \begin{minipage}{0.45\textwidth}
 \includegraphics[width=5.5cm,angle=-90]{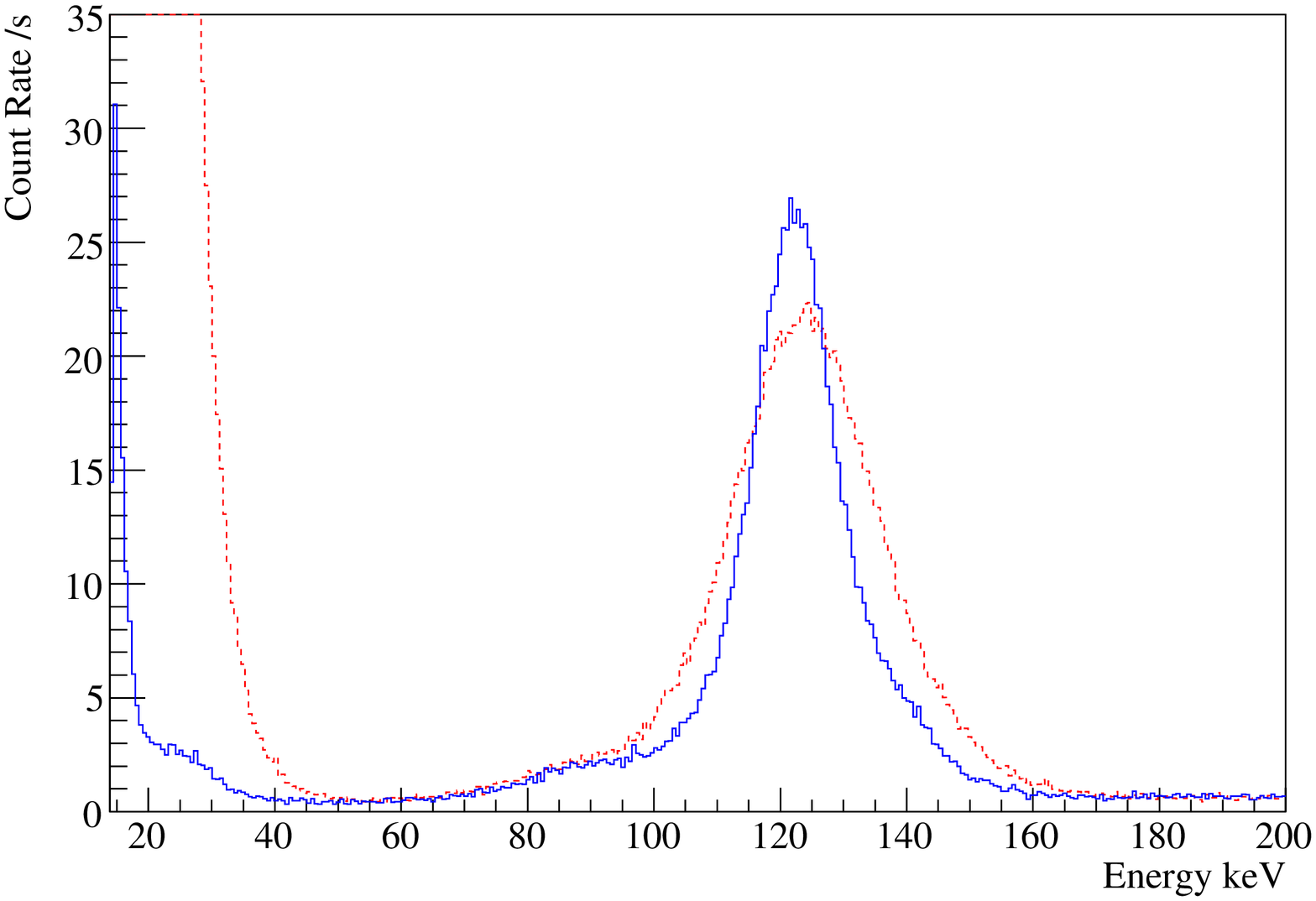}
  \end{minipage}%
 \hspace{0.1cm}
  \begin{minipage}{0.45\textwidth}
\includegraphics[width=5.5cm,angle=-90]{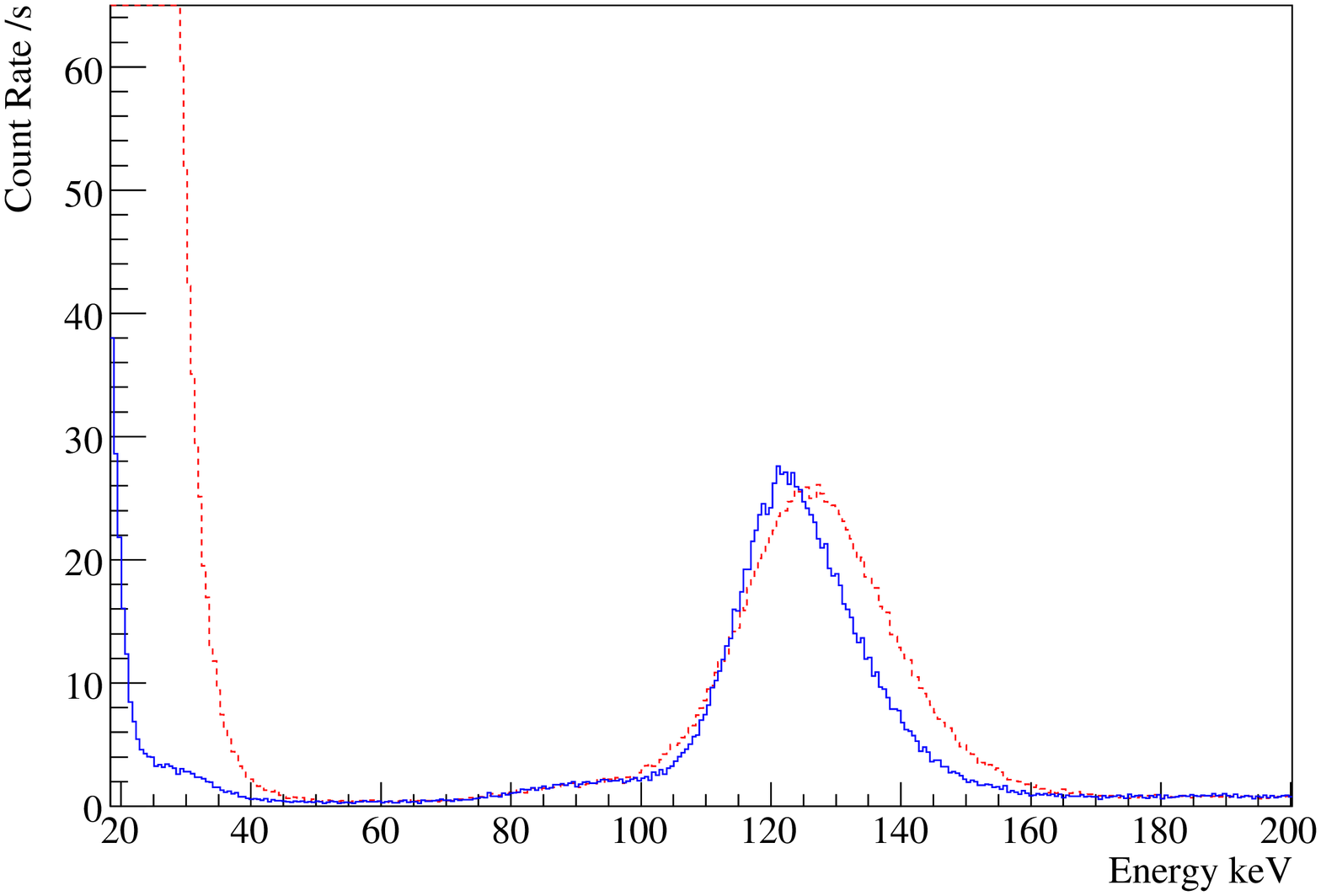}

  \end{minipage}%
\caption{\label{fig:co57} $^{57}$Co spectrum of crystal A (left) and
B (right) at room temperature (23$^{\circ}$C)(dashed line) and under
cooling (5$^{\circ}$C) (solid line).}
\end{figure}

Low energy spectra show a significant enhancement in resolution under
cooling.  Figures \ref{fig:am241} and \ref{fig:co57} show the spectral
change observed for $^{241}$Am and $^{57}$Co sources cooled to 5$^\circ$C .  A clear
improvement is observed with cooling; the photopeaks become narrower such that
other spectral features become visible. In addition, the low energy
threshold drops such that features at 30 keV become apparent.

\begin{figure}
\begin{center}
\includegraphics[width=12cm,height=7cm]{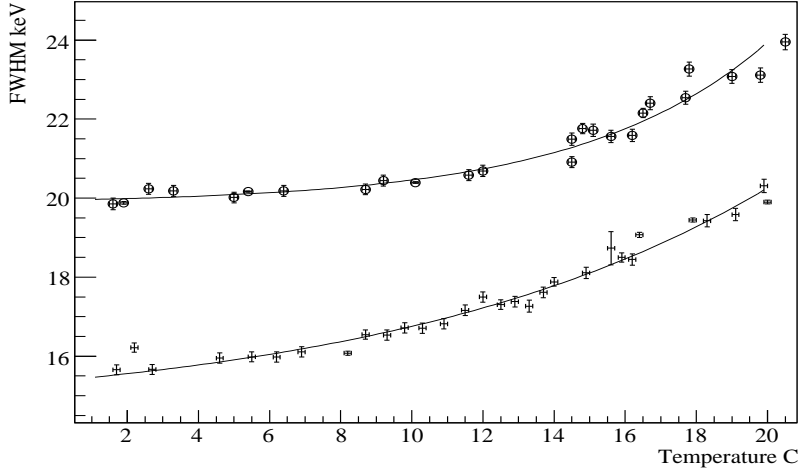}
\caption{\small Change in the FWHM of the 122 keV line of $^{57}$Co as a
  function of temperature for crystals A(circles) and B(points).  Fits
  to A: (19.84$\pm$0.03)+(0.10$\pm$0.01)e$^{(0.185\pm0.009) T}$ and B(14.6$\pm$0.2)+(0.8$\pm$0.1) e$^{(0.097\pm0.007) T}$.}
\label{fig:122linewithtemp}
\end{center}
\end{figure}

\begin{figure}
\begin{center}
\includegraphics[width=12cm,height=7cm]{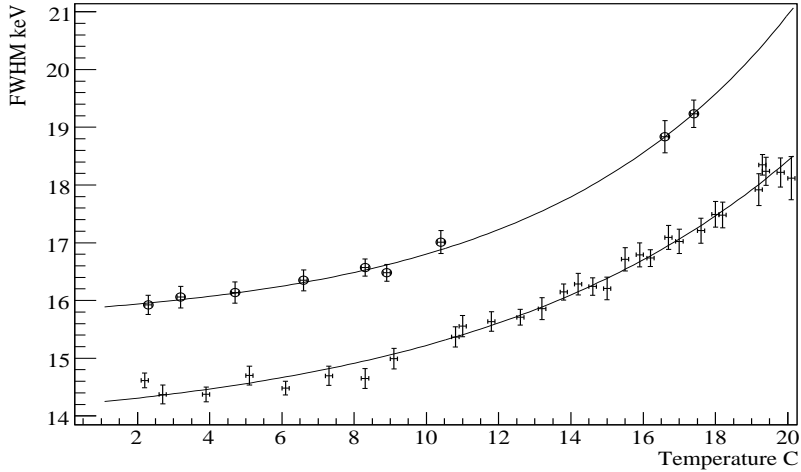}
\caption{\small Change in the FWHM of the 60 keV line of $^{241}$Am as a
  function of temperature for crystals A(circles) and B(points).  Fits
  to A: (15.5$\pm$0.3)+(0.3$\pm$0.2) e$^{(0.15\pm0.04) T}$ and B (13.7$\pm$0.2)+(0.5$\pm$0.1)e$^{(0.11\pm0.01) T}$}

\label{fig:60linewithtemp}
\end{center}
\end{figure}

Measurements of the FWHM of the 122 keV line from $^{57}$Co and the 60
keV line from $^{241}$Am were made
as a function of temperature for the two crystals, and are shown in
Figures  \ref{fig:122linewithtemp} and  \ref{fig:60linewithtemp}.  The sources were not collimated and the cathodes of both detectors were equally illuminated.  Both
data sets are fitted by a single exponential function combined with a
constant offset.  The exponential form of the fitted function
represents the magnitude and response of the crystal to temperature.  Supporting this, the resulting fit parameters of the 60keV and 122keV
lines are compatible for each individual crystal, but not compatible
between crystals i.e. the temperature response is a detector property.  

The leakage current is reduced such that it is insignificant compared to other sources of photopeak broadening.  For crystal A, the leakage current component of the FWHM is $lc(T)_A=(0.10\pm0.01)e^{(0.185\pm0.009) T}$ and for crystal B $lc_(T)_B=(0.8\pm0.1) e^{(0.097\pm0.007) T}$.

The residual FWHM comes from other components unaffected by temperature such as electronic noise, cable length, detector performance, and any geometrical broadening.

Cooling to 5$^\circ$C brings a 25$\%$ improvement to the
resolution of the 122 keV line on crystal A and 22$\%$ improvement on
crystal B.  For these crystals further cooling will not bring significant improvements to the energy resolution.
\subsection{Low Energy Thresholds}
\label{thresholds}
\begin{figure}
  \begin{minipage}{0.45\textwidth}
 \includegraphics[width=7 cm,,height=7cm,angle=0]{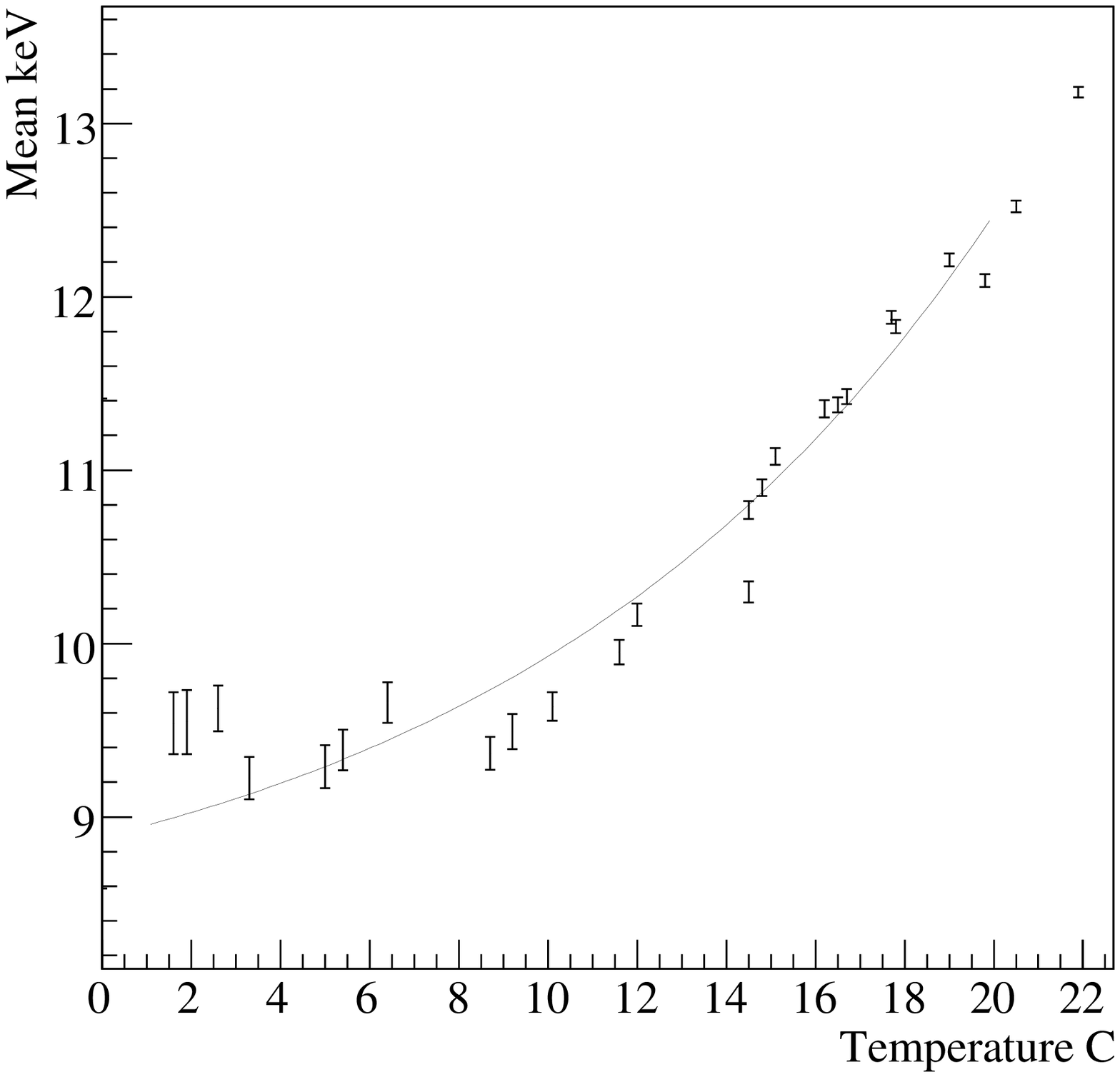}
  \end{minipage}%
 \hspace{0.5cm}
  \begin{minipage}{0.45\textwidth}
\includegraphics[width=7 cm,,height=7cm,angle=0]{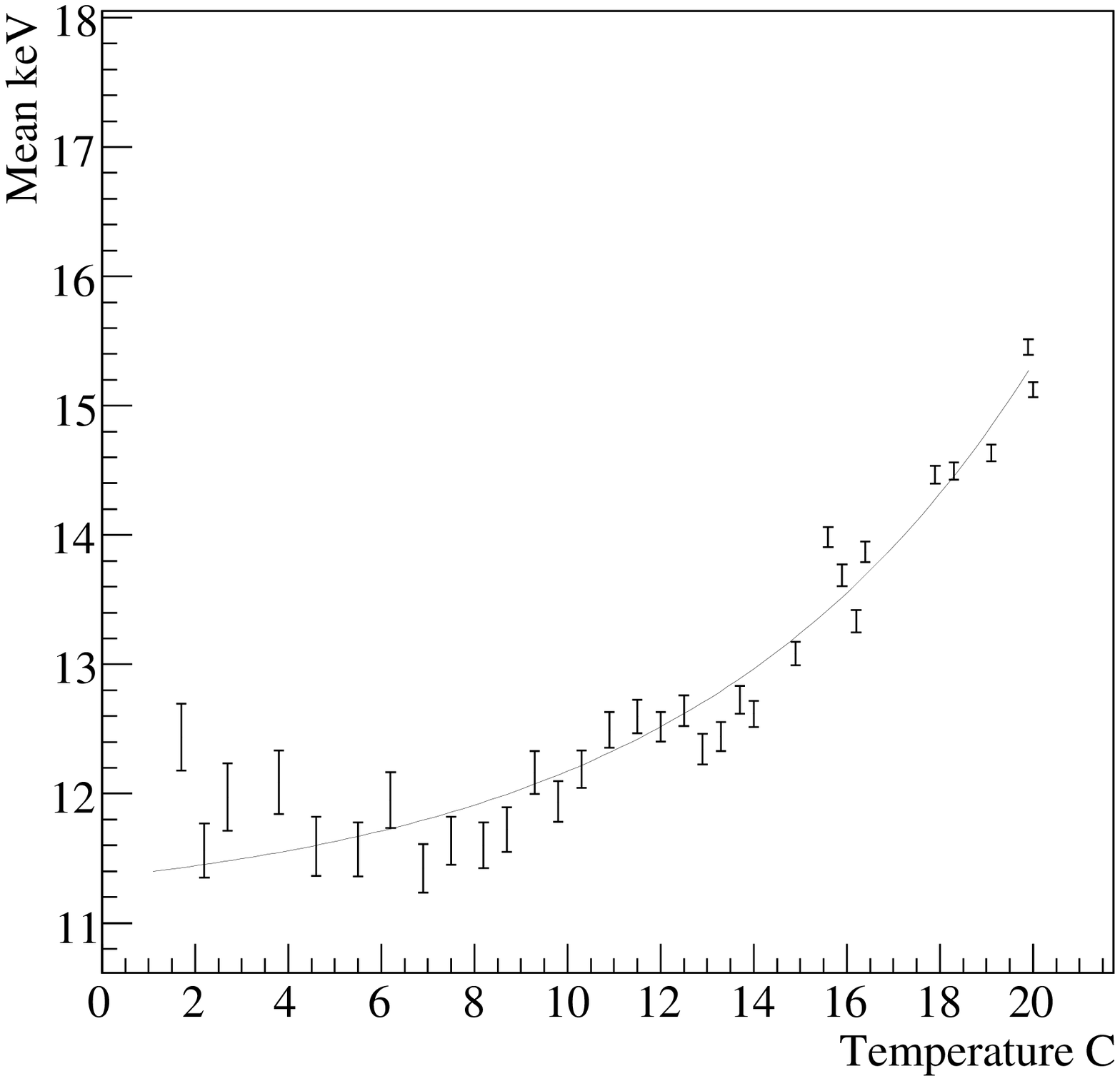}
  \end{minipage}%
\caption{\label{fig:threshold} Variation of the Poisson mean of the pedestal as a function of temperature from crystal A (left) and B(right).}
\end{figure}

\begin{figure}
  \begin{minipage}{0.45\textwidth}
 \includegraphics[width=7 cm,height=7cm,angle=0]{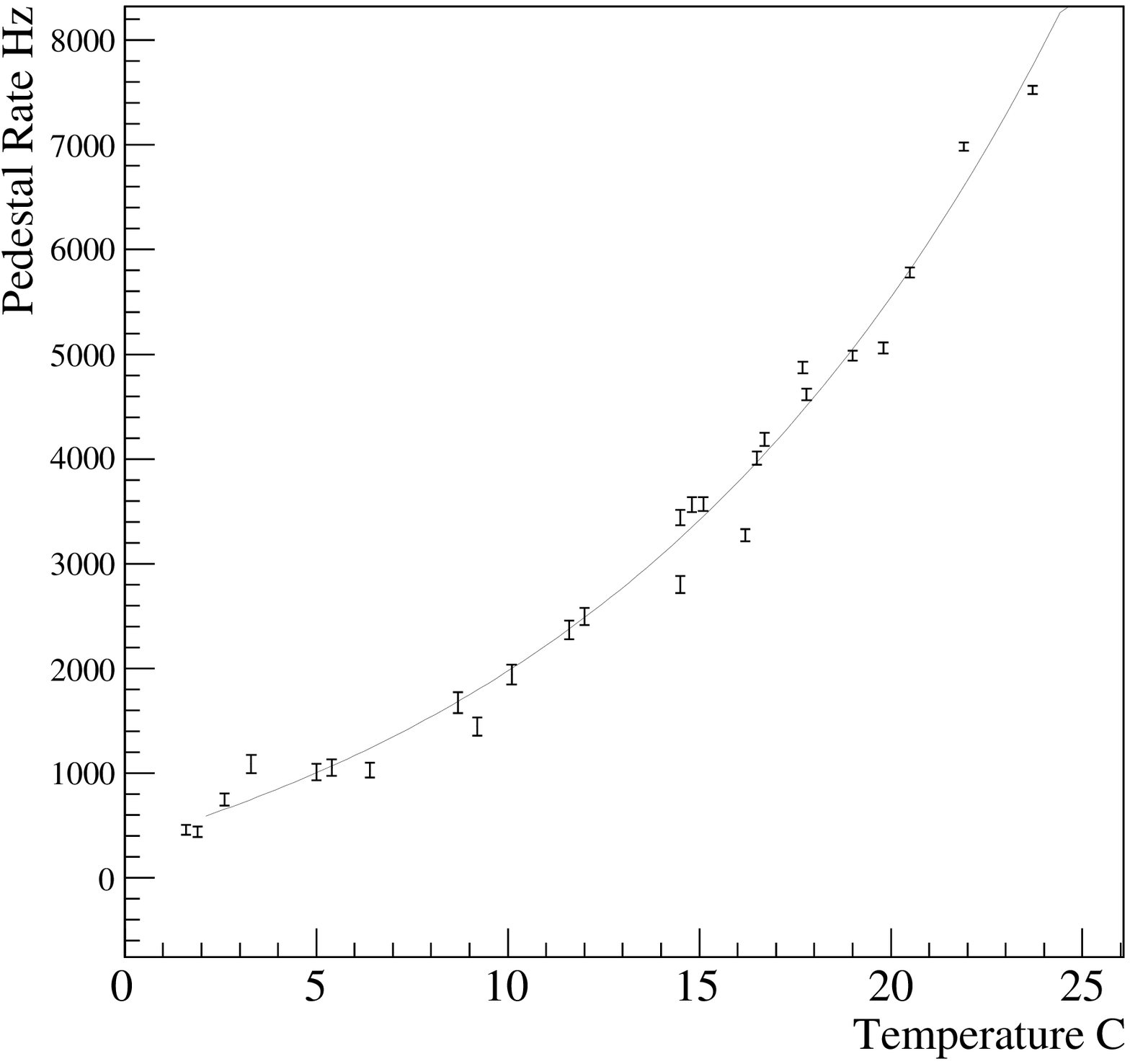}
  \end{minipage}%
 \hspace{0.5cm}
  \begin{minipage}{0.45\textwidth}
\includegraphics[width=7 cm,height=7cm, angle=0]{{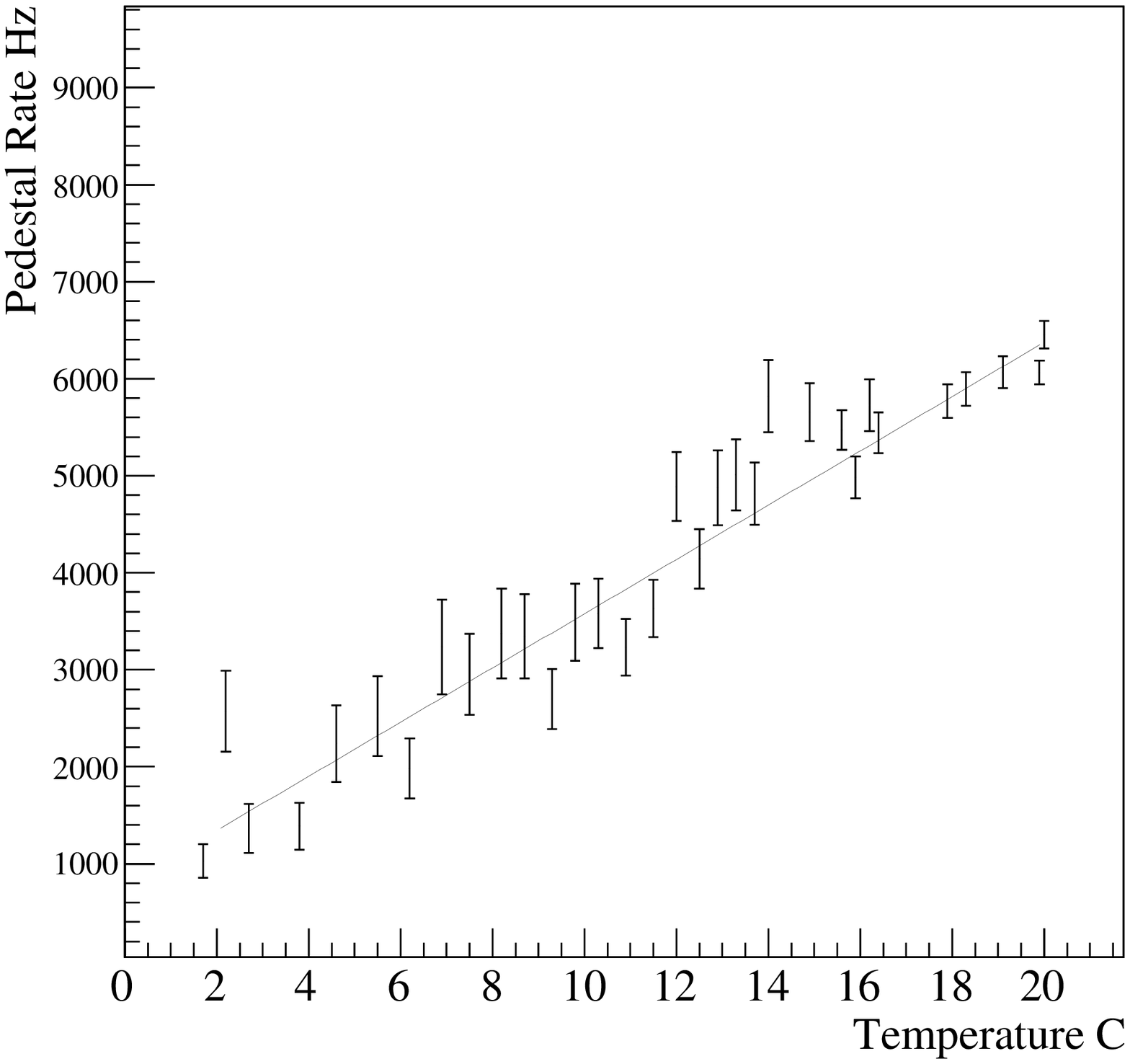}}
  \end{minipage}%
\caption{\label{fig:threshold_num} Variation of the number pedestal counts as a
  function of temperature from crystal A (left) and B(right).}
\end{figure}

Cooling the CZT crystals results in a significant improvement in lower
energy threshold due to the reduction in leakage current, with the pedestal feature observed in the spectra
clearly reducing in size.  Poisson fits to the pedestal were made for
spectra from both crystals for different temperatures. Figure
\ref{fig:threshold} shows how the fitted mean reduces with
temperature.  Additionally the height of the noise pedestal also
diminishes as shown in Figure \ref{fig:threshold_num}. Similar to the
photopeak resolutions, the energy threshold reaches a minimum value at
$\sim$5$^{\circ}$C and further cooling does not improve the threshold.

In rare search mode, the low energy threshold is always far greater
than that observed during calibrations with a high rate source.  The
low energy threshold can be thought of as the energy at which the noise
rate drops below the real physics event rate.  Low rate searches
therefore experience higher low energy thresholds.  Whilst it is
possible to observe low energy lines $\sim$30 keV during calibrations,
the low energy threshold is often  $\sim$50 keV or higher in low
background operation.

The energy threshold is therefore defined to be the energy at which
the noise rate falls below a pre-determined value.  Using the results
of the Poisson fit to the noise pedestal (shown in Figures
\ref{fig:threshold} and \ref{fig:threshold_num}) for a particular
operating temperature, one can find the energy at which the noise rate
falls below a given value. For example, for a desired threshold event
rate of 10$^{-4}$ Hz at 23$^{\circ}$C crystal A would have a low
energy threshold of 37.6 keV.  Lowering the temperature to  $5^{\circ}$C, would reduce this
threshold to 29.1 keV.  Similarly for crystal B, the low energy
threshold would be 41.8 keV at room temperature reducing to 33.8 keV
with cooling.

\section{Resolution Function}
\begin{figure}
\begin{center}
\includegraphics[width=12cm,height=7cm]{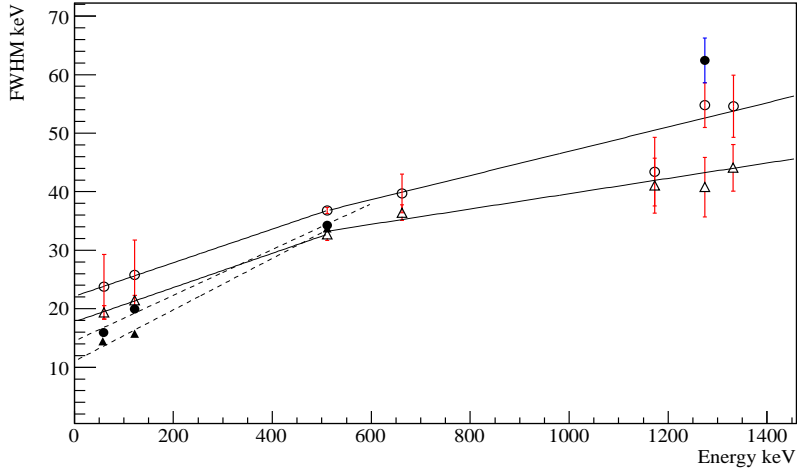}
\caption{\small FWHM versus energy for crystals A (circles) and B
  (triangles), open and filled markers denote room temperature and
  cooled data respectively. The solid lines show the trend from the 511 keV line and above. Below this the 511 keV line the trends change, solid represents the room temperature results and the dashed line is with maximum cooling (no leakage current).}
\label{fig:resolutionAB}
\end{center}
\end{figure}

Figure \ref{fig:resolutionAB} shows the resolution functions observed
with the two crystals A and B.  Since no spectral change was observed
for these two crystals for the higher energy lines (511 keV and
above), the two straight lines above the 511 keV point represent the
trend for both room temperature and under cooling.  For the lower
energy lines, the trends differ and the same linear trend is no longer
followed. This illustrates how temperature affects the resolution
function. This also shows how determining the resolution function
using high energy lines and extrapolating to lower energies may result
in a wrong determination of the widths of the low energy lines. 

The behaviour of the resolution of the two COBRA crystals is different to
that of the commercial detector, as shown in Figure\ref{fig:resolutionfunction}, where the room temperature and cooled
resolution functions are well separated and virtually parallel. In
the COBRA case there is no discernable difference between the room
temperature and cooled resolution functions at 511 keV.  Here the
derived resolution functions merge with FWHMs of $\sim$ 35 keV.  This
resolution is only reached for the highest energy lines observed with
the commercial detector and we assume that for higher energy lines the
resolution functions would also begin to merge.  

For better-performing commerical
detector the leakage current is a major source of broadening across
the range 60 to 1300 keV.  For the COBRA detectors, the leakage
current component is less important with respect to the the other
sources of broadening such as the electronic noise.  For the COBRA
detectors the electronic noise term is higher and therefore the
leakage current is less significant.  For energies above $\sim$300 keV
the leakage current is insignificant with respect to the other sources
of broadening. Therefore to improve the resolutions of higher energy
lines, methods to counteract this other component must be found.

\label{resolution_function}

\section{Conclusion}
\label{conclusion}
A systematic study of the behaviour of two CZT Co-Planar Grid crystals
under moderate cooling has been made. Significant improvements in
photopeak resolution and low energy thresholds were observed.  A
temperature of only 5$^{\circ}$C was found to be sufficient, below
this no further improvement was observed.

The change in FWHM of photopeaks as a function of temperature was
found to be well fitted by an exponential trend. For each crystal, the 60 keV and 122 keV datasets were
fitted to this exponential function and the resulting fit parameters were
found to be compatible, i.e. the behaviour of the photopeak resolution
with temperature is the same for all energies. This is interpreted as leakage current decreasing with decreasing temperature.  

Both crystals exhibit $\sim$21$\%$ and $\sim$16$\%$ improvement in
resolution at 60 keV and 122 keV respectively under cooling.
Improvements at higher energies above $\sim$300 keV are
indiscernable, due to the insignificance of leakage current broadening
at these energies.  The resolution here is thought to be dominated by
geometric and electronic effects. 
Typically leakage currents are found to contribute $\sim$5 keV
to the widths of photopeaks.  For these two crystals, other noise sources which are not affected by the temperature of the crystal, such as the electronic noise, contribute a large part of the FWHM of the photopeaks. It is clear that the reduction of this contribution will
be beneficial.

The resolution functions, trend of FWHM with photopeak energy, have been determined for the commercial eV PRODUCTs detector and the two COBRA detectors.  They both show linear increases in FWHM as a function of energy. This is interpreted as being a geometric effect, where energy depositions on one side of the detector give systematically large signals than the opposing side.  Under cooling, the commercial detector shows improvements from 60 to 1300 keV.  For the highest energy lines observed with the commercial detector the FWHMs are $\approx$35 keV.  With the COBRA detectors this FWHM is reached at a lower energy at 511 keV.  Above this there is no improvement seen with the cooling as the leakage current is not the dominant source of broadening.

Moderate cooling reduces the low energy threshold due to the reduction
in leakage current.  We consider a low
background threshold event rate of 10$^{-4}$ Hz, such that the COBRA
experiment operates at, and show that cooling would reduce the
experimental threshold of these detectors by $\sim$20$\%$, to 30 and
34 keV for crystals A and B.  We consider that cooling the crystals to
5$^{\circ}$C of the already running COBRA experiment would bring about
similar improvements. This would allow a search for the second-forbidden
unique electron capture of $^{123}$Te, peak at 30.5 keV.  

For the present COBRA experiment, mild cooling will improve the physics search sensitivity to low energy lines only and will bring no improvements to the search for high energy lines.  However, the improvements in low energy threshold increase the physics reach of the experiment.

To improve the resolution on the high energy lines, such as $^{116}$Cd
2.8 MeV line, work should focus on investigating and reducing the
broadening effects which are dominant at this energy.  This currently
appears to be due to the design of the CPG grids and the pre-amplifier
electronics.  Future work will also concentrate on reducing the electronic noise, with the aim of pushing the low background threshold to well below 30 keV.

\end{document}